# Towards Affective Drone Swarms: A Preliminary Crowd-Sourced Study

Truong-Huy D. Nguyen, Kasper Grispino, Damian Lyons
Department of Computer and Information Science, Fordham University, New York, NY
{tnguyen88, kgrispino, dlyons}@fordham.edu

## Abstract

Drone swarms are teams of autonomous unmanned aerial vehicles that act as a collective entity. We are interested in humanizing drone swarms, equipping them with the ability to emotionally affect human users through their non-verbal motions. Inspired by recent findings in how observers are emotionally touched by watching dance moves, we investigate the questions of whether and how coordinated drone swarms' motions can achieve emotive impacts on general audience. Our preliminary study on Amazon Mechanical Turk led to a number of interesting findings, including both promising results and challenges.

## 1 Introduction

Human-to-human interactions are filled with affects and emotions, which are conveyed and perceived through both verbal and non-verbal communication. In order to humanize AI-powered technologies, accounting for emotions in the interactions with human users is essential. Our research focuses on non-verbal communication of robots that can invoke emotional responses similar to that with human.

Recently, psychologists found that observers' feelings vary when watching dance moves of different dynamic characteristics [1]. Specifically, it was found that round (curvy) motions are correlated to more positive feelings than edgy ones, motion energy is correlated with valence, and impressive movements enhance positive responses. Such findings intrigued and drove robotic researchers to investigate how to construct robots whose emotions can invoke similar emotions.

There have been efforts investigating how to encode emotions in solo drones' motions [2], [3]. Cauchard et al. [2] detailed a design process to encode personalities in drones' flight paths, so that viewers can infer their personalities from the way they move. They map characteristics of personalities to dimensions of drone movements such as altitude, direction, speed, angles (toward the user), and reaction time and compliance to basic commands. In a similar vein, Szafir et al. [3] devised a formalism defining a set of flight motion primitives, the combination of which form flightpaths capable of communicating intents to viewers.

Building upon related works in encoding emotions in single drones' motions, we are investigating mechanisms to construct drone swarms' motions that can elicit emotions the same way people would feel when watching dance moves. Since drone swarms consist of more than one member, they are capable of enacting coordinated moves that cannot be executed by solo drones. This is the main motivation behind our research, aiming to capitalize on such additional degrees of freedom. In the preliminary study reported in this paper, we are examining the effects of drone swarms' motion edginess on three emotional aspects: felt, perceived emotions, and safeness.

The organization of the paper is as follows. We will first lay out some recent related works, before detailing the motion parameter we will study and report. Finally, we describe the preliminary crowd-sourced study set up and present our findings.

## 2 Related Works

Many related works are concerned with using aerial or ground robot movements to depict emotions. Most of them however focus on directing solo-drone flight paths rather than multi-drones.

### 2.1 Encoding personality and intents with drones

Cauchard et al. [2] investigated the problem of using drones' flight paths to reflect their predefined personality, with the goal to convey such personality to viewers. The researchers started with folklore stereotypes of personalities, as depicted in the seven dwarves in Walt Disney's movie Snow White. Next, workshops with designers were conducted to identified descriptive vocabulary associated with each personality. Finally, they map such vocabulary on to dimensions of drone movement parameters such as altitude, direction, speed, angles (toward the user), and reaction time and compliance to basic commands. In-lab validation with human subjects (N=20) shows the effectiveness of the resultant encoding scheme.

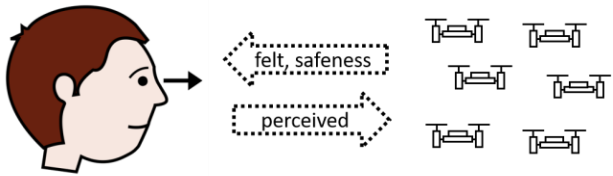

Figure 1. Types of emotions as invoked when watching drone swarms' motions

Szafir et al. [3] tackled the problem of defining flight motion primitives, the combination of which form flightpaths capable of communicating intents to viewers. They identified a total of 11 such primitives, categorized into two types: core and interactive. Core primitives are motions essential to movements such as takeoff, landing, cruise, etc., while interactive are those revolving around a specific target such as approaching, avoiding, and departing movements. Robot intents are specified in terms of where they will move to next, e.g., to the right, to the left, or stop in front of object, etc. Their evaluation demonstrated that adopting motion primitives allows the drone's intent to be inferred easier from viewers.

## 2.2 Measures of Affects

In this work, we would like to gain some insights on three aspects of emotion perception invoked through watching drone swarm motions: (1) **perceived** emotion, (2) **felt** emotion, and (3) **safeness**. As depicted in Figure 1, perceived emotion refers to the emotion viewers attribute to the robot group, while felt emotion and safeness refer to the emotions invoked internally on the viewer's part.

To measure the perceived and felt emotions, we are using a modified version of the PANAS scale [4], the details of which will be described in the Study section. The PANAS scales were originally created and tested for the purpose of reporting ones' affect feelings. It comprises of 20 descriptive words representing positive and negative feelings, 10 for positive and 10 for negative. Sample positive words include *interested*, *excited*, *strong*, *enthusiastic*, *proud*, *active*, *attentive*, *inspired*, etc., while negative ones include *afraid*, *irritable*, *scared*, *distressed*, *ashamed*, etc. Adopting PANAS scale to measure affect level, typically participants would be asked to rate, on 5-point Likert scales, their feelings.

The list of words in PANAS scale were reported to sometimes be too long or ambiguous, so there have been efforts to devise shorter forms of PANAS that are more succinct and clearer while maintaining equal effectiveness in estimating affects [5], [6]. For instance, the I-PANAS-SF [6] is a shorter version of PANAS, which comprises of 5 negative and 5 positive words, and is friendly to even international participants. It was shown that affects deduced from I-PANAS-SF scales are well correlated with those from the original PANAS scales.

## 3 Targeted Motion Parameters

We drew inspirations from previous research findings on how observers feel when watching dance moves [1] or ground robots' motions [7].

Christensen et al. [1] found that watching dance movements can elicit affective responses from observers, with movement roundedness correlating with positive emotions, while edgy ones negative. As such, edginess was identified to be an important motion parameter in imparting emotions. Similarly, Saerbeck and Bartneck [7] conducted a study to investigate how people perceived, i.e., attribute emotions to, a ground robot (Roomba[1]) when observing its motions. In the study, they found that acceleration and curvature have a significant effect on the perceived affective state.

In our research, we are interested in validating motion-based motifs based on four motion parameters, namely Volume, Velocity, Acceleration, and Edginess
- Volume refers to the space which the drone swarm occupy during their operation
- Velocity refers to the average speed at which the drones are moving
- Acceleration refers to the change in velocity that can be described qualitatively as jerkiness.
- Edginess refers to the acuteness of direction change angles. In one extreme (high edginess), the drones change their direction movement at 45 degrees, while in the other extreme (low edginess), the drones change direction through curvy paths.

In addition, we would also like to survey observers on their perceptions to understand when and how drone swarms can be perceived as safe/unsafe. To establish successful interactions with robot agents, trust or feeling of safety is an important factor [8].

In this paper, we are reporting our study in validating the effect of motion edginess on affects, given that the performers are a drone swarm. Specifically, to study the effects of edginess, we compare the motions shown in Figure 2, in which two drones make the horizontal 8 figures (similar to the infinity symbol) with different settings for edginess: one is the ordinary curvy 8 motion, the other one a boxy motion, as if the flightpath is comprised of two squares placing next to one another.

---
[1] http://www.irobot.com/For-the-Home/Vacuuming/Roomba.aspx

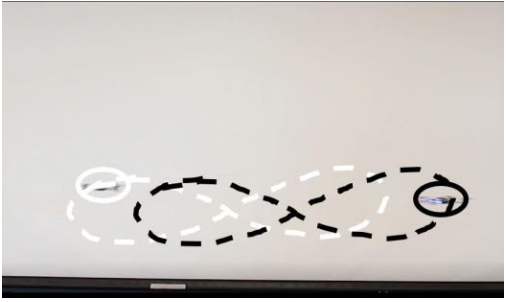 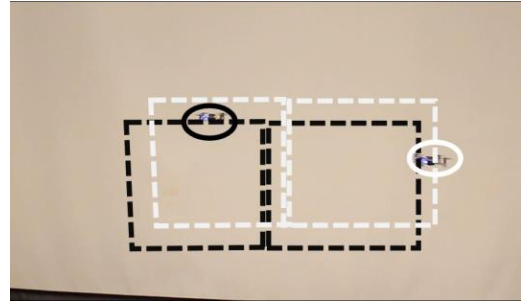

(a) Circle-8  (b) Square-8

Figure 2. The drone swarm's motions, with the dotted flightpaths showing the shapes of the motions, and solid eclipses the drones. Each video clip comprises of two drones flying in sync to keep the collective velocity zero, with (a) showing motions in Circle-8 shape, and (b) motions in Square-8 shape.

## 5 Experiment Setup

In our preliminary study, we would like to study how edginess affects emotions.

### 5.1 Implementation of drone swarm behaviors

We implemented the framework on BitCraze's CrazyFlie 2.0 [9]. CrazyFlie is a palm-sized micro-drone that was designed for lightweight and versatile maneuver. We recorded two clips of drone swarm motions (Figure 2)

- Circle-8, which takes the shape of a horizontal 8 shape
- Square-8, which is similar to Circle-8, except that the cyclic flight path is boxy, instead of curvy

We chose these two motions for our study since they have the exact same shape but different degrees of edginess: Circle-8 is smooth with no acute change of direction, while Square-8 has frequent 90-degree direction turns.

### 5.2 Measuring affects

We extended the I-PANAS-SF survey [6] to allow participants more freedom to express the reported affective state. Besides the original 10 words to be rated on 5-point Likert scales, two modifications were made to the survey. First, we added an additional word named "Others" to the list of emotion words to be rated. This allows participant to describe in their own words what emotion they feel appropriate and its scale. Second, participants are asked to justify/explain their ratings in free text. This will help shed light on what aspects of the motions may have led the to the reported emotions.

### 5.3 Study Setup

Participants are recruited from Amazon Mechanical Turk [10] for our repeated measure design with one independent variable, i.e., edginess (edgy - curvy).

Participants are asked to watch two clips of different configurations of edginess, i.e., Circle-8 and Square-8. After watching the clips, they will rate their perceived and felt emotions using our modified I-PANAS-SF surveys. Besides they are asked about the perceived safeness when hypothetically interacting with the drone swarms.

## 6 Results and Discussions

We recruited a total of 24 MTurk workers, 12 watching Circle-8 first, and 12 Square-8 first. The randomization of clip order to eliminate the sequence effect of their exposure to drone swarms' motions.

### 6.1 Quantitative results

Figure 3 depicts the results as obtained from the study, plotted with boxplots the perceived and felt emotions on the Circle- and Square-8 motions, while Figure 4 that of the felt safeness. On average, the perceived affects of Circle-8 and Square-8 are rated similarly (Figure 3 left column) at slightly greater than 10 for positivity, and about 3 for negative[2]. The felt emotions of both types, on the other hand, are rated at about 10 for positive, and 1 for negative. The felt safeness for both motions are neutral[3], with mean values at about 2.2 (Figure 4).

---

[2] The affect values (positive or negative) obtained from I-PANAS-SF range between 0 and 20, the higher the more significant.

[3] Felt safeness is a single rating, so its value ranges from 0 to 4.

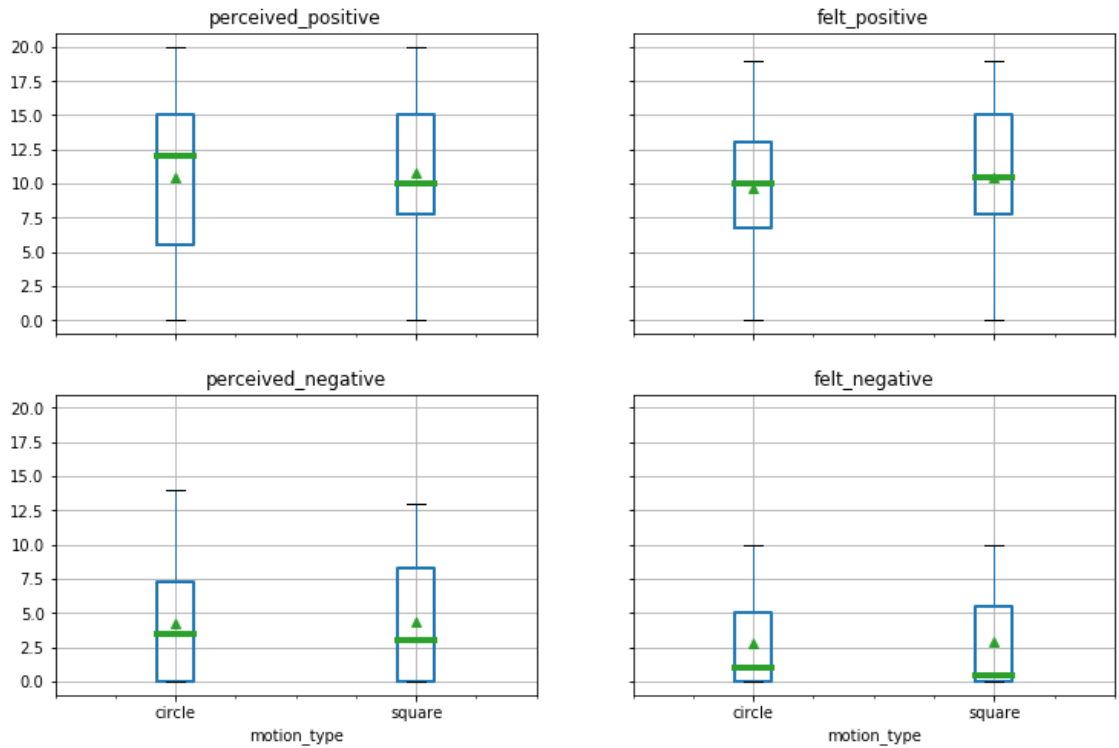

Figure 3. Boxplots showing perceived and felt emotion ratings of Circle- and Square-8 motions. The horizontal bar in each box indicates the median value, while the small triangle represents the mean value.

Our first observation is that the edginess of the motions (circle or square) does not appear to yield significant impact on either type of emotions (i.e., perceived and felt), as their average ratings are very similar. This suggests that the choice of edginess as implemented, i.e., curvy versus 90-degree direction change, may not be significant enough to make a notable difference. That said, the motions seem to impart more positivity than negativity. This is an encouraging result, indicating that drone swarms' motions may be used to invoke emotions of positive valence. We will need to examine edginess in tandem with motion shapes to reliably draw any conclusion on their effects on felt or perceived emotions.

### 6.2 Qualitative results

Overall, perceived emotions were invoked slightly more strongly than felt emotions, as observed in the plots. This means that observers would associate motions as a means for the drone swarm to express their own internal affective state. More specifically, many observers reported neutral felt feelings after watching the clips, using statements such as:

- "*I can't say I'm having any kind of emotion.*"
- "*I was not effected by them much. Just standard drone activity.*"
- "*it was rather relaxing, but for the most part they did not change the way I generally am feeling*"
- "*watching the drones didn't change my emotions*"
- etc.

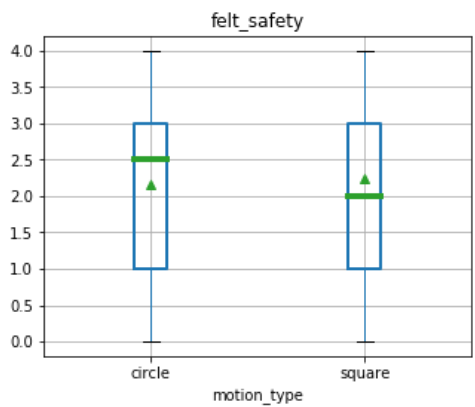

Figure 4. Boxplots showing felt safeness ratings of Circle- and Square-8 motions. The horizontal bar in each box indicates the median value, while the small triangle represents the mean value.

The safeness ratings, as shown in Figure 4 being neutral and leaning a bit towards more safeness, indicates some reservation from the participants. As we examined the explanations provided to justify the ratings, we found that safeness ratings are mostly associated with either the perception of **control**, drones' **size**, or that they are just "**toys**". We believe the perception that drones are toys are closely related to whether one can be in control of them or not. For example,

- "Very trained people controlling the drones. I would feel safe" (rating 3/4)
- "I own a drone and these seemed to be in control" (rating 3/4)
- "They are small and look harmless" (rating 4/4)
- "They were not erratic" (rating 3/4)
- "I think that to operate these are a safe action." (rating 3/4)
- "consumer drones are like toys with blades on them, how safe do you think they can be" (rating 1/4)
- "they seemed harmless. I felt like they were toys." (rating 2/4)

### 6.3 Unexpected Responses and Implications

The qualitative questions in the survey, in which participants can respond with free-form text, gave us a few unexpected insights.

First, the **synchronization** and **coordination** aspects of the motions caught significant attention from some participants. For example, on justifying their ratings on perceived emotions of the Circle-8, one participant said,
- "*The movements are soft and well-coordinated, it looks like a couple dancing while they are talking each other in a sweet way*".

On the Square-8 motion, some stated
- "*They seemed to be trying to keep in sync in a happy way,*" and
- "*At first, I felt they were fighting, but as time went on, it became much more apparent that they were in sync with each other. It was more like a dance than a fight. I felt almost an attraction, like 2 people dancing the tango.*"

In this case, the participants identified the swarm's motions as those of two drones trying to interact with one another in an intimate manner. Similarly, another participant just noticed the synchronization, declaring "I don't think they're feeling much at all, they're flying in a very coordinated fashion." As such, we suspect that:
- Drones might be used for acting in dramatic sketches and their motions may be authored to mimic acting moves with emotional motifs.
- To ensure that drones are perceived as a single collective entity, we need to increase the number of drones in the swarm to a sufficient level that masks away their individuality. Otherwise, they may be tracked and identified as individuals.

Second, as revealed in the safeness ratings, the perception of safeness can be associated with how the drones appear to be **in control** of their motions. Unlike ground robots, air-borne ones such as drones are sensitive to unavoidable changes in environment conditions such as air flows and turbulence, even when they are just hovering at one place. As such, to humanize AI-powered drones, we believe that the perception of drones being in control is something that should not be overlooked.

## 7 Conclusions

Multi-drone swarms possess higher degrees of freedom, as compared to solo-drones, with the potential to manipulate volumes, shapes, dynamicity (e.g., velocity, acceleration, and motion edginess), as well as the transitions between them. In this work, we examined how edginess affect viewers' felt, perceived emotions, and feelings of safeness. The results of our crowd-sourced perception studies show that in drone swarms, coordination is an important factor to get judged by observers. Besides, with drones that move in sync with one another, they would be perceived more positively than negatively.

As future work, we would like to first increase the scale of the study to a larger number of participants and more comprehensively analyze the effects of motion parameters using techniques such as ANOVA to identify statistically significant effects. Next, we will expand the study by formalizing some motion motifs based on a broadened set of motion parameters such as velocity, acceleration, and space. These motifs can form a repertoire of building blocks or vocabulary upon which drone swarms' motions can be authored for communicating messages and emotions to viewers. This has application in situations where drones are tasked to assist human, such as search and rescue missions [11].

Another direction of future work is to investigate how the framework can be adapted to account for higher levels of interaction. As of now, there is no interaction between observers and drone swarms; perception is merely through observing. Interactions in reciprocal manners would add more complexity to the framework, as the feeling of safety, and subsequently trust, would need to be gained from the observers before natural interactions can be achieved.

## References

[1] J. F. Christensen, F. E. Pollick, A. Lambrechts, and A. Gomila, "Affective responses to dance," *Acta Psychol. (Amst).*, vol. 168, pp. 91–105, Jul. 2016.

[2] J. R. Cauchard, K. Y. Zhai, M. Spadafora, and J. A. Landay, "Emotion encoding in Human-Drone Interaction," in *2016 11th ACM/IEEE International Conference on Human-Robot Interaction (HRI)*, 2016, pp. 263–270.

[3] D. Szafir, B. Mutlu, and T. Fong, "Communication of intent in assistive free flyers," in *Proceedings of the 2014 ACM/IEEE international conference on Human-robot interaction - HRI '14*, 2014, pp. 358–365.

[4] D. Watson, L. A. Clark, and A. Tellegen, "Development and validation of brief measures of positive and negative affect: the PANAS scales.," *J. Pers. Soc. Psychol.*, vol. 54, no. 6, pp. 1063–70, Jun. 1988.

[5] K. Kercher, "Assessing Subjective Well-Being in the Old-Old," *Res. Aging*, vol. 14, no. 2, pp. 131–168, 1992.

[6] E. R. Thompson, "Development and validation of an internationally reliable short-form of the Positive and Negative Affect Schedule (PANAS)," *J. Cross. Cult. Psychol.*, vol. 38, no. 2, pp. 227–242, 2007.

[7] M. Saerbeck and C. Bartneck, "Perception of affect elicited by robot motion," in *Proceeding of the 5th ACM/IEEE international conference on Human-robot interaction - HRI '10*, 2010, p. 53.

[8] J. D. Lee and K. A. See, "Trust in Automation: Designing for Appropriate Reliance," *Hum. Factors J. Hum. Factors Ergon. Soc.*, vol. 46, no. 1, pp. 50–80, 2004.

[9] Bitcraze AB, "Crazyflie 2.0," 2016. [Online]. Available: https://www.bitcraze.io/crazyflie-2/. [Accessed: 12-Feb-2018].

[10] M. Buhrmester, T. Kwang, and S. D. Gosling, "Amazon's Mechanical Turk," *Perspect. Psychol. Sci.*, vol. 6, no. 1, pp. 3–5, Jan. 2011.

[11] V. Kumar, "Equipping Drones for Search and Rescue.," *Pract. Sail.*, vol. 38, pp. 24–27, 2012.